\documentclass[longauth]{aa} 
\usepackage{graphicx}
\usepackage{txfonts}
\def\lsw{1}
\def\tuorla{2}
\def\athensUniv{3}
\def\bonnMPIfR{4}
\def\inafTorino{5}
\def\kentuckyUniv{6}
\def\lulinAI{7}
\def\lulinDP{8}
\def\perugiaUniv{9}
\def\crimeanSAI{10}
\def\stpetersburgUniv{11}
\def\corkCIT{12}
\def\ulughbeg{13}
\def\isaacnewton{14}
\def\metsahovi{15}
\def\masarykUniv{16}
\def\abastumani{17}
\def\potsdamAI{18}
\def\trebur{19}
\def\coyote{20}
\def\alaskaUniv{21}
\def\victoriaUniv{22}
\def\hiroshima{23}
\def\stlouis{24}
\def\michiganUniv{25}
\def\astron{26}
\def\amsterdamUniv{27}
\def\iram{28}
\def\arizonaUniv{29}

\begin{document}

   \title{Testing the inverse-Compton catastrophe scenario \\ in the intra-day
          variable blazar S5 0716+71}

   \subtitle{I. Simultaneous broadband observations\thanks{Partially based on observations 
             obtained with INTEGRAL, an ESA project with instruments and
             science data centre funded by ESA member states
             (especially the PI countries: Denmark, France, Germany, Italy,
             Switzerland, Spain, Czech Republic and Poland), and with the participation
             of Russia and the USA. 
             Partially based on observations by the Whole Earth Blazar Telescope (WEBT); 
             for questions regarding the availability of the data from the WEBT campaign, 
             please contact M. Villata ({\tt villata@to.astro.it}).
             Partially based on observations collected at the German-Spanish
             Astronomical Center, Calar Alto, operated by the Max-Planck-Institut f\"ur 
             Astronomie, Heidelberg, jointly with the Spanish National Commission for 
             Astronomy.
             Partially based on observations with the 100-m telescope of the MPIfR 
             (Max-Planck-Institut f\"ur Radioastronomie) in Effelsberg, Germany.} 
             during November 2003}

   \author{L. Ostorero \inst{\lsw,\tuorla}
          \and S. J. Wagner \inst{\lsw}  
          \and J. Gracia \inst{\athensUniv}     
          \and E. Ferrero \inst{\lsw}    
          \and T. P. Krichbaum \inst{\bonnMPIfR} 
	  \and S. Britzen \inst{\bonnMPIfR}       
	  \and A. Witzel \inst{\bonnMPIfR}       
          \and K. Nilsson \inst{\tuorla}       
          \and M. Villata \inst{\inafTorino}     
          \and U. Bach \inst{\inafTorino,\bonnMPIfR}   
          \and D. Barnaby   \inst{\kentuckyUniv} 
	  \and S. Bernhart \inst{\bonnMPIfR}      
          \and M. T. Carini \inst{\kentuckyUniv}  
          \and C. W. Chen \inst{\lulinAI}    
          \and W. P. Chen \inst{\lulinAI,\lulinDP}     
          \and S. Ciprini \inst{\tuorla,\perugiaUniv} 
          \and S. Crapanzano \inst{\inafTorino}  
          \and V. Doroshenko \inst{\crimeanSAI}  
          \and N. V. Efimova \inst{\stpetersburgUniv}  
          \and D. Emmanoulopoulos \inst{\lsw} 
          \and L. Fuhrmann \inst{\perugiaUniv,\inafTorino,\bonnMPIfR}      
	  \and K. Gabanyi \inst{\bonnMPIfR}      
          \and A. Giltinan \inst{\corkCIT}
          \and V. Hagen-Thorn \inst{\stpetersburgUniv} 
          \and M. Hauser \inst{\lsw}     
          \and J. Heidt \inst{\lsw}     
          \and A. S. Hojaev \inst{\lulinAI,\ulughbeg,\isaacnewton}    
          \and T. Hovatta \inst{\metsahovi}               
          \and F. Hroch \inst{\masarykUniv}  
          \and M. Ibrahimov \inst{\ulughbeg}   
	  \and V. Impellizzeri \inst{\bonnMPIfR} 
          \and R. Z. Ivanidze \inst{\abastumani}
          \and D. Kachel \inst{\lsw}     
	  \and A. Kraus \inst{\bonnMPIfR} 
          \and O. Kurtanidze \inst{\abastumani,\lsw,\potsdamAI}
          \and A. L\"ahteenm\"aki \inst{\metsahovi}  
          \and L. Lanteri \inst{\inafTorino}    
          \and V. M. Larionov \inst{\stpetersburgUniv}
          \and Z. Y. Lin \inst{\lulinAI}      
          \and E. Lindfors \inst{\tuorla}     
          \and F. Munz \inst{\masarykUniv}      
          \and M. G. Nikolashvili \inst{\abastumani}
          \and G. Nucciarelli \inst{\perugiaUniv} 
          \and A. O'Connor \inst{\corkCIT}
          \and J. Ohlert \inst{\trebur}     
          \and M. Pasanen \inst{\tuorla}      
          \and C. Pullen \inst{\coyote}  
          \and C. M. Raiteri \inst{\inafTorino}  
          \and T. A. Rector \inst{\alaskaUniv}  
          \and R. Robb \inst{\victoriaUniv}   
          \and L. A. Sigua \inst{\abastumani}  
          \and A. Sillanp\"a\"a  \inst{\tuorla} 
          \and L. Sixtova \inst{\masarykUniv}   
          \and N. Smith \inst{\corkCIT}       
          \and P. Strub \inst{\lsw}      
          \and S. Takahashi \inst{\hiroshima}   
          \and L. O. Takalo \inst{\tuorla}  
          \and C. Tapken \inst{\lsw}     
          \and J. Tartar \inst{\stlouis}     
          \and M. Tornikoski \inst{\metsahovi}      
          \and G. Tosti \inst{\perugiaUniv}      
          \and M. Tr\"oller \inst{\metsahovi}      
          \and R. Walters   \inst{\kentuckyUniv} 
          \and B. A. Wilking \inst{\stlouis}  
          \and W. Wills     \inst{\kentuckyUniv} 
	  \and I. Agudo \inst{\bonnMPIfR}         
          \and H. D. Aller \inst{\michiganUniv}       
          \and M. F. Aller \inst{\michiganUniv}        
	  \and E. Angelakis \inst{\bonnMPIfR}     
	  \and J. Klare \inst{\bonnMPIfR}       
	  \and E. K\"ording \inst{\bonnMPIfR}    
	  \and R. G. Strom \inst{\astron,\amsterdamUniv}    
          \and H. Ter\"asranta \inst{\metsahovi}  
	  \and H. Ungerechts \inst{\iram}   
	  \and B. Vila-Vilar\'o \inst{\arizonaUniv}   
   }

   \offprints{L.Ostorero@lsw.uni-heidelberg.de}

   \institute{Landessternwarte Heidelberg-K\"onigstuhl, K\"onigstuhl, 69117
              Heidelberg, Germany
\and Tuorla Observatory, University of Turku, V\"ais\"al\"antie 20, 21500 Piikki\"o, Finland 
\and Section of Astrophysics, Astronomy \& Mechanics, Department of
     Physics, University of Athens, Panepistimiopolis, 157 84 Zografos, Athens,
     Greece
\and Max-Planck-Institut f\"ur Radioastronomie, Auf dem H\"ugel 69, 53121 Bonn, Germany 
\and Istituto Nazionale di Astrofisica (INAF), Osservatorio Astronomico di Torino,
     via Osservatorio 20, 10025 Pino Torinese (TO), Italy 
\and Department of Physics and Astronomy, Western Kentucky University, 1 Big Red
     Way, Bowling Green, KY 42104, USA 
\and Institute of Astronomy, National Central University, 300 Jungda Road,
     Jungli City 320-54, Taoyuan, Taiwan, ROC
\and Department of Physics, National Central University, 300 Jungda Road,
     Jungli City 320-54, Taoyuan, Taiwan, ROC
\and Osservatorio Astronomico, Universit\`a di Perugia, Via B. Bonfigli, I-06126 Perugia, Italy
\and Crimean Laboratory of the Sternberg Astronomical Institute, University of
     Moscow, Russia; P/O Nauchny, 98409 Crimea, Ukraine
\and Astronomical Institute, St. Petersburg State University, Universitetsky
     pr. 28, Petrodvoretz, 198504 St. Petersburg, Russia
\and Cork Institute of Technology, Dept. of Applied Physics \& Instrumentation, 
     Rossa Avenue, Bishoptown, Cork, Ireland 
\and Ulugh Beg Astronomical Institute, Center for Space Research, Uzbek
     Academy of Sciences, Astronomicheskaya 33, 700052 Tashkent, Uzbekistan
\and Isaac Newton Institute, Uzbekistan Branch, Astronomicheskaya 33, 700052 Tashkent, Uzbekistan
\and Mets\"ahovi Radio Observatory, Helsinki University of Technology,
     Mets\"ahovintie 114, 02540 Kylm\"al\"a, Finland 
\and Institute of Theoretical Physics and Astrophysics, Faculty of Science, Masaryk University, 
     Kotl\'arsk\'a 2, 611 37 Brno, Czech Republic 
\and Abastumani Astrophysical Observatory, 383762 Abastumani, Georgia
\and Astrophysikalisches Institut Potsdam, An der Sternwarte 16, 14482 Potsdam, Germany
\and Michael Adrian Observatory, Astronomie Stiftung Trebur, Fichtenstrasse 7,
     65468 Trebur, Germany
\and Coyote Hill Observatory, P.O. Box 930, Wilton, CA 95693, USA
\and Department of Physics and Astronomy, University of Alaska Anchorage, 3211
     Providence Dr., Anchorage, AK 99508, USA
\and Department of Physics and Astronomy, University of Victoria, BC, Canada
\and Faculty of Information Sciences, Hiroshima-City University, 3-4-1,
     Ozuka-Higashi, Asa-minami-ku, Hiroshima, 731-3194, Japan
\and Department of Physics and Astronomy, University of Missouri--St. Louis, 8001
     Natural Bridge Road, St. Louis, MO 63121
\and Department of Astronomy, Dennison Building, University of Michigan, Ann Arbor,
     MI 48109 USA
\and ASTRON, Postbus 2, 7990 AA Dwingeloo, The Netherlands
\and Astronomical Institute, University of Amsterdam, Kruislaan 403, 1098 SJ
     Amsterdam, The Netherlands
\and IRAM, Avd. Div. Pastora 7NC, 18012 Granada, Spain
\and University of Arizona, Steward Observatory, 933 N. Cherry Ave., Tucson, AZ 85721, USA
          }

\date{Received...; accepted...}

\titlerunning{Testing the Inverse-Compton catastrophe scenario in S5 0716+71}
     
\authorrunning{L.\ Ostorero et al.}                   

\abstract
{Some intra-day variable, compact extra-galactic radio sources show brightness
temperatures severely exceeding $10^{12}$ K, the limit set by catastrophic 
inverse-Compton (IC) cooling in sources of incoherent synchrotron
radiation.
The violation of the IC limit, actually possible under non-stationary 
conditions, would lead to IC avalanches in the soft-$\gamma$--ray energy band 
during transient periods.}
{For the first time, broadband signatures of possible IC catastrophes were 
searched for in a prototypical source, S5 0716+71.}
{A multifrequency observing campaign targetting S5 0716+71 was
carried out during November 06--20, 2003.
The observations, organized under the framework of the European Network for the
Investigation of Galactic nuclei through Multifrequency Analysis
(ENIGMA) together with a campaign by the Whole Earth Blazar 
Telescope (WEBT), involved a pointing by the soft-$\gamma$--ray
satellite INTEGRAL, optical, near-infrared, sub-millimeter,
millimeter, radio, as well as Very Long Baseline Array (VLBA) monitoring.}
{S5 0716+71 was very bright at radio frequencies and in
a rather faint optical state ($R =14.17-13.64$) during the
INTEGRAL pointing; significant inter-day and low intra-day variability 
was recorded in the radio regime, while typical fast variability
features were observed in the optical band.
No obvious correlation was found between the radio and optical emission. 
The source was not detected by INTEGRAL, neither by the X-ray monitor JEM-X 
nor by the $\gamma$-ray imager ISGRI, but upper limits to the source emission 
in the 3--200 keV energy band were estimated.
A brightness temperature $T_{\mathrm b} >2.1\times 10^{14}$ K (violating the IC 
limit) was inferred from the variability observed in the radio regime, but no 
corresponding signatures of IC avalanches were recorded at higher energies.}
{In the most plausible scenario of negligible contribution of the interstellar
scintillation to the observed radio variability, the absence of the signatures of IC catastrophes 
provides either a lower limit $\delta \ga 8$ to the Doppler factor affecting the
radio emission or strong constraints for modelling of the Compton-catastrophe
scenario in S5 0716+71.}


\maketitle


\section{Introduction}

The phenomenon of intra-day variability (IDV; Wagner \& Witzel
\cite{wagner1995}) has been a long-standing problem since its disco\-very 
in the late 1960s (see e.g.\ Racine et al. \cite{racine1970}; 
Witzel et al. \cite{witzel1986}; Heeschen et al. \cite{heeschen1987}).
The occurrence of IDV appears to be more common in flat-spectrum
extragalactic sources dominated by a very compact core in Very Long Baseline 
Interferometry (VLBI) maps
(Quirrenbach et al. \cite{quirrenbach1992}, \cite{quirrenbach2000}).
With recent Very Long Baseline Array (VLBA) measurements, 
Kovalev et al. (\cite{kovalev2005}) showed that IDV sources typically exhibit
a higher compactness and core-dominance on sub-milliarcsecond scales than
non-IDV ones; they also found that a higher amplitude of intraday variations
characterizes sources with a higher flux density in an unresolved VLBA
component, and that the most variable sources tend to have the most compact structure.

Rapid variations in flux may be caused by mechanisms intrinsic to the source.
In this case, causality arguments 
would imply that the variability originates from
very compact regions of the AGN, thus characterized by high photon
densities and brightness temperatures.
In sources of incoherent synchrotron radiation, very high photon densities 
would lead to catastrophic cooling via inverse-Compton (IC) scattering
of the synchrotron radiation by the high-energy electrons, with a 
production of high-energy radiation much higher than observed 
(Hoyle et al. \cite{hoyle1966}). 
Kellermann \& Pauliny-Toth (\cite{kellermann1969})
showed that the onset of catastrophic radiation losses, occurring when the 
photon energy density in the emission region exceeds the energy density of 
the magnetic field, limits the maximum observed brightness temperature
to the so-called `IC limit' $T_{\mathrm b} \la 10^{12}$ K 
(see also Readhead \cite{readhead1994}).
They also noticed, however, that this limit may be significantly 
exceeded under non-stationary conditions.
Evidence of severe violations of the IC limit was reported for many compact 
IDV radio sources (see e.g.\ Quirrenbach et al. \cite{quirrenbach1989}; Wagner
\& Witzel \cite{wagner1995}; Kedziora-Chudczer et al. \cite{kedziora1997}),
although no information about corresponding IC-scattered emission 
is available for any of them.

Alternatively, propagation effects, like variations of the absorption along the line of
sight, deflection of the light in the potential well of foreground  
stars (microlensing; Chang \& Refsdal \cite{chang1979}), and, 
in the radio regime, 
interstellar scintillation (ISS; Rickett \cite{rickett1990}; 
Rickett et al. \cite{rickett1995}), are possible 
extrinsic-source mechanisms which have been invoked to explain IDV.
On the one hand, microlensing was shown to be an unlikely explanation of the rapid
variability in the best-studied IDV sources, due to the frequency of the
flaring activity,
the statistical asymmetry of the light curves, the short time scales of
variability, and the non-zero lag measured between optical and radio variations
(see Wagner \cite{wagner1992} and references therein). 
On the other hand, 
ISS is not expected to play a major
role at millimeter and sub-millimeter wavelengths, the transition
between weak and strong scintillation regimes occurring
in the centimeter domain (Rickett et al. \cite{rickett1995}).

In order to reconcile intrinsic variations with the theoretical limit,
different explanations, like beaming of the emission 
due to bulk relativistic motion (Rees \cite{rees1967}), 
coherent radiation mechanisms 
(Baker et al. \cite{baker1988}; Benford \cite{benford1992};
Lesch \& Pohl \cite{lesch_pohl1992}), 
propagation of a shock in an underlying, stable relativistic jet 
(Qian et al. \cite{qian1991}), were proposed. 
Neither the broad-band nature of the spectral energy distribution (SED) observed
in IDV sources (see e.g.\ Quirrenbach et al. \cite{quirrenbach1989}; Wagner et
al. \cite{wagner1993}, \cite{wagner1996}) nor the VLBI observations 
(see e.g.\ Gabuzda \cite{gabuzda2000})
strongly support any of them, leaving the alternative explanation that the IC
limit might actually be violated and inverse-Compton catastrophes occur 
during transient periods 
(Kellermann \& Pauliny-Toth\cite{kellermann1969}; Slysh \cite{slysh1992}).
In this paper, this hypothesis is explored for a prototypical source, S5 0716+71.

This source is one of the brightest and best-studied BL Lacertae objects in the
sky. It was one of the prime targets for investigating the mechanism responsible for IDV,
and the first source in which simultaneous variations in
the radio and optical bands, indicating a possible intrinsic 
origin of the observed IDV, were reported (Wagner et al. \cite {wagner1990};
Quirrenbach et al. \cite {quirrenbach1991}).
Moreover, it exhibited IDV during all the past optical studies 
and almost all the radio campaigns carried 
out during the last two decades 
(Heeschen et al. \cite{heeschen1987}; 
Wagner et al. \cite {wagner1990}, \cite {wagner1996};
Heidt \& Wagner \cite{heidt1996};
Ghisellini et al. \cite{ghisellini1997};
Sagar et al. \cite{sagar1999};
Quirrenbach et al. \cite{quirrenbach2000};
Villata et al. \cite{villata2000}; 
Nesci et al. \cite{nesci2002};
Kraus et al. \cite{kraus2003};
Raiteri et al. \cite{raiteri2003}).
The IDV duty cycle of the source (the fraction of time in which the object is
variable) derived from these studies is $\sim 90$\%.

Deep maps of the source obtained with the Very Large Array (VLA) show a core-halo structure 
on the arcsecond scale.
VLBI observations over more than 20 years at centimeter wavelengths reveal a very
compact source, with evidence of a core-dominated jet
structure extending several tens of milliarcseconds to the North
(Eckart et al. \cite{eckart1986}, \cite{eckart1987}; 
Witzel et al. \cite{witzel1988}; Polatidis et al. \cite{polatidis1995};
Jorstad et al. \cite{jorstad2001}). The milliarcsecond jet is misaligned with
respect to the VLA jet by $\sim 75^{\circ}$ (Britzen et
al. \cite{britzen2005}; see also Eckart et al. \cite{eckart1987}).

Controversial scenarios, involving a wide range of proper motions (0.05--1.2
mas/year), were proposed for the kinematics of the S5 0716+71 jet components,
the more recent ones poin\-ting towards apparent velocities which are atypically
fast for BL Lac objects (see Bach et al. \cite{bach2005}, and references therein).

The redshift of the source is still unknown, although the starlike appearance and the
absence of any signature of a host galaxy in deep images had set a lower limit
of $z>0.3$ (Quirrenbach et al. \cite{quirrenbach1991}; Stickel et al. \cite{stickel1993};
Wagner et al. \cite{wagner1996}), which will be used throughout this paper. 
More recently, Sbarufatti et al. (\cite{sbarufatti2005})
suggested a higher lower limit of $z>0.52$.
The exact brightness temperature of the source can therefore not be
determined: 
lower limits up to $T_{\mathrm b}\sim10^{17}$~K 
were inferred from radio IDV at 5 GHz 
(Quirrenbach et al. \cite{quirrenbach1991}; Wagner et al. \cite{wagner1996}),
whereas a limit of 
$T_{{\mathrm b},\, z=0}=1.85\times 10^{13}$ K was estimated from the 
constraints on the core size derived through interferometric measurements 
performed in August 2003 with the Very Long Baseline Array (VLBA) 
at 15 GHz (Kovalev et al. \cite{kovalev2005}).
This source is hence an ideal target for the investigation of radiative
signatures of IC ca\-tastro\-phes.

S5 0716+71 has been detected at GeV energies with a steep $\gamma$-ray spectrum  
(Hartman et al. \cite{hartman1999}), but the soft-$\gamma$--ray part of its spectral 
energy distribution is poorly known: upper li\-mits to the emission were
provided by OSSE (McNaron-Brown et al. \cite{mcnaron1995}) and COMPTEL 
(Sch\"onfelder et al. \cite{schonfelder2000}); a recent reanalysis of the
COMPTEL data (Collmar \cite{collmar2006}) yielded a source detection in
the 3--10 MeV energy range.

Synchrotron-self-Compton (SSC) emission models, which can reproduce the 
GeV $\gamma$-ray emission of the source (see e.g. Ghisellini et
al. \cite{ghisellini1997}), predict the SED peak to occur in the MeV--GeV
domain. 
The MeV detection reported by Collmar (\cite{collmar2006})
is indeed consistent with this scenario.
The above SSC modelling involves a wide range of Lorentz factors of the 
radiating particles, up to $\gamma\sim 10^5$, in order to explain the 
high-energy component of the spectrum.

In a Compton catastrophe, the main role is expected to be played by electrons
responsible for the bulk of the radio emission in the GHz regime,
where the violations of the IC limit have been observed.
Assuming an observed peak frequency of the radio spectrum 
of a few GHz, and a magnetic field 
$B\ga 10^{-3}/\delta$ $\mu$G 
($\delta$ being the Doppler factor of the emitting region), 
the above electrons would be characterized by Lorentz factors not greater 
than a few times $10^2$.
First-order IC scattering by these particles would generate bursts of radiation at 
$\sim~10^{14}-10^{15}$ Hz, which might blend with the synchrotron radiation
produced by the high-energy tail of the same distribution of particles.
The dominant loss-term would be second-order IC scattering 
(see e.g.\ Kellermann \& Pauliny-Toth \cite{kellermann1969}; 
Bloom \& Marscher \cite{bloom1996}), also responsible for the onset of the 
ca\-tastro\-phe, and would boost photons into the $10^{18}-10^{20}$ Hz frequency range.
Flares of IC-scattered radiation should therefore be observed in this energy
range whenever the Compton limit is violated.
The efficient cooling associated with a Compton ca\-tastro\-phe 
would then rapidly restore the actual brightness temperature of the source.

The advent of the soft-$\gamma$--ray INTEGRAL satellite offers an 
unprecedented chance of investigating this effect.

A multifrequency campaign involving an INTEGRAL pointing of the source and 
simultaneous radio, millimeter, sub-millimeter, near-infrared, and
optical photometric monitoring, as well as VLBA observations, 
was organized and carried out during November 2003.
For the first time, constraints on the brightness temperature and on the 
IC emission of the source were tested simultaneously against each other.

The first results of the ground-based observing campaign 
are presented in Sect. \ref{sec_ground} of this paper; the INTEGRAL observations are
described in Sect.\ \ref{sec_integral}; the simultaneous spectral energy distribution of
the source is presented in Sect.\ \ref{sec_sed}; in Sect.\ \ref{sec_conclusions}
we discuss our results and draw conclusions.

\section{Ground-based multifrequency observations}
\label{sec_ground}
The multiwavelength observations, organized under the framework of 
the ENIGMA\footnote{{\tt http://www.lsw.uni-heidelberg.de/projects/enigma/}} 
collaboration together with a 
WEBT\footnote{{\tt http://www.to.astro.it/blazars/webt/}\\
(see e.g.\ Mattox et al. \cite{mattox1998}; 
Villata et al. \cite{villata2004}, 
Raiteri et al. \cite{raiteri2005}, and references therein)}
campaign, were scheduled for the period November 06--20, 2003 (hereafter
referred to as the core campaign) to provide the low-energy counterpart of the 
INTEGRAL observation of S5 0716+71. 
Since the source underwent an unprecedented outburst phase at radio and millimeter 
frequencies du\-ring September--October 2003, 
the campaign was started earlier (October 2003). 
When the core campaign began on November 06 (JD = 2452949.5),
S5 0716+71 was still brighter in the radio and millimeter domain
than in all past studies.
A rather faint optical state ($R \sim 14.17-13.64$) characterized the source
in this band for the whole duration of the INTEGRAL pointing. 
During the last two days of the core campaign the source entered a brighte\-ning
phase, and the ground-based monitoring was continued till May 2004. 

We here present the results of the core campaign in terms of 
radio light curves at 32 and 37 GHz, and an optical $R$-band light curve.
Flux variation ranges at other radio, millimeter and sub-millimeter
frequencies are included in the simultaneous SED presented in Sect.\ \ref{sec_sed};
the corresponding light curves will be discussed in Agudo et al. (\cite{agudo2006})
and Fuhrmann et al. (in prep.).
The ground-based observing facilities whose contribution to the core campaign 
is presented in this paper 
are listed in Table~\ref{table:telescopes}.

\subsection{Radio data}
\label{sec_radiodata}

\begin{table*}
\caption{List of the ground-based observing facilities whose contribution to the core
         campaign is presented in this paper in the form of 
         light curves and/or flux variation ranges in the SED.
         a: Radio, millimeter and submillimeter antennas, listed in
         ascending order of the lowest observing frequency. 
         b: Optical telescopes, listed in order of longitude, and their
         contribution to the core campaign in terms of $R$-band useful
         data (and observing nights).
         The offsets indicate the corrections applied to the
         data sets in order to eliminate instrumental inconsistencies (when a
         range is given, different offsets were used on different nights).}
\label{table:telescopes}      
\centering          
\begin{tabular}{llccc}    
\hline
\hline
\multicolumn{4}{c}{{\bf a. Radio, millimeter and sub-millimeter observatories}}\\
\hline                  
Observatory & Location        &  Telescope diameter   & \multicolumn{2}{c}{Observing frequencies}\\    
(telescope) & ~~              &   (m)                 & \multicolumn{2}{c}{(GHz)}\\    
\hline
Westerbork  & The Netherlands &  14  $\times$ 25      & \multicolumn{2}{c}{1.392, 1.67}\\
UMRAO       & Michigan, USA   &  26                   & \multicolumn{2}{c}{4.8, 8.0, 14.5}\\
Effelsberg  & Germany         &  100                  & \multicolumn{2}{c}{4.85, 10.45, 32.0}\\
Mets\"ahovi & Finland         &  13.7                 & \multicolumn{2}{c}{22.2, 36.8}\\
IRAM        & Spain           &  30                   & \multicolumn{2}{c}{86.0, 230.0}\\
HHT-SMTO    & Arizona, USA    &  10                   & \multicolumn{2}{c}{345.0}\\
JAC (JCMT)  & Hawaii          &  15                   & \multicolumn{2}{c}{350.0, 664.0}\\              
\hline                  
\hline
\multicolumn{4}{c}{{\bf b. Optical observatories}}\\
\hline                  
Observatory & Location &  Telescope diameter   & $R$-band data          & Offsets\\    
~~          & ~~       &  (cm)                 & (nights)               & ~~\\    
\hline
Lulin                        & Taiwan                 &  100           & 153 (7) &  0.\\
Mt. Maidanak                 & Uzbekistan             &  150           &   2 (1) &  0. \\
Abastumani                   & Georgia                &  70            & 655 (6) &  [$-$0.03;0.]\\
Crimean                      & Ukraine                &  70            &   3 (2) &  +0.03\\
Tuorla                       & Finland                &  103           & 102 (2) &  [+0.01;+0.02]\\
MonteBoo                     & Czech Republic         &  62            &  59 (2) &  [+0.05;+0.07]\\
Perugia                      & Italy                  &  40            &   7 (4) &  0.\\
Heidelberg                   & Germany                &  70            & 333 (8) &  [+0.03;+0.06]\\
Michael Adrian               & Germany                &  120           &  90 (4) &  +0.03 \\
Torino                       & Italy                  &  105           &  19 (3) &  $-$0.02\\
Hoher List                   & Germany                &  106           &  41 (2) &  +0.14\\
Calar Alto                   & Spain                  &  220           & 374 (3) &  [+0.01;+0.02]\\
Roque de los Muchachos (KVA) & Spain                  &  35            &  58 (5) &  $-$0.01\\
Roque de los Muchachos (WHT) & Spain                  &  420           & 751 (1) &  0.\\ 
Bell                         & Kentucky, USA          &  60            &   3 (1) & 0.\\
St. Louis                    & Missouri, USA          &  36            &   3 (1) & 0.\\
Kitt Peak (WIYN)             & Arizona, USA           &  90            &  21 (6) & 0.\\
Coyote Hill                  & California, USA        &  28            &  73 (1) & 0.\\
University of Victoria       & Canada                 &  50            & 102 (1) & 0.\\
\hline                  
\end{tabular}
\end{table*}

\begin{figure*}
\mbox{\hskip -2.5truecm\includegraphics[width=22.cm,height=14cm]{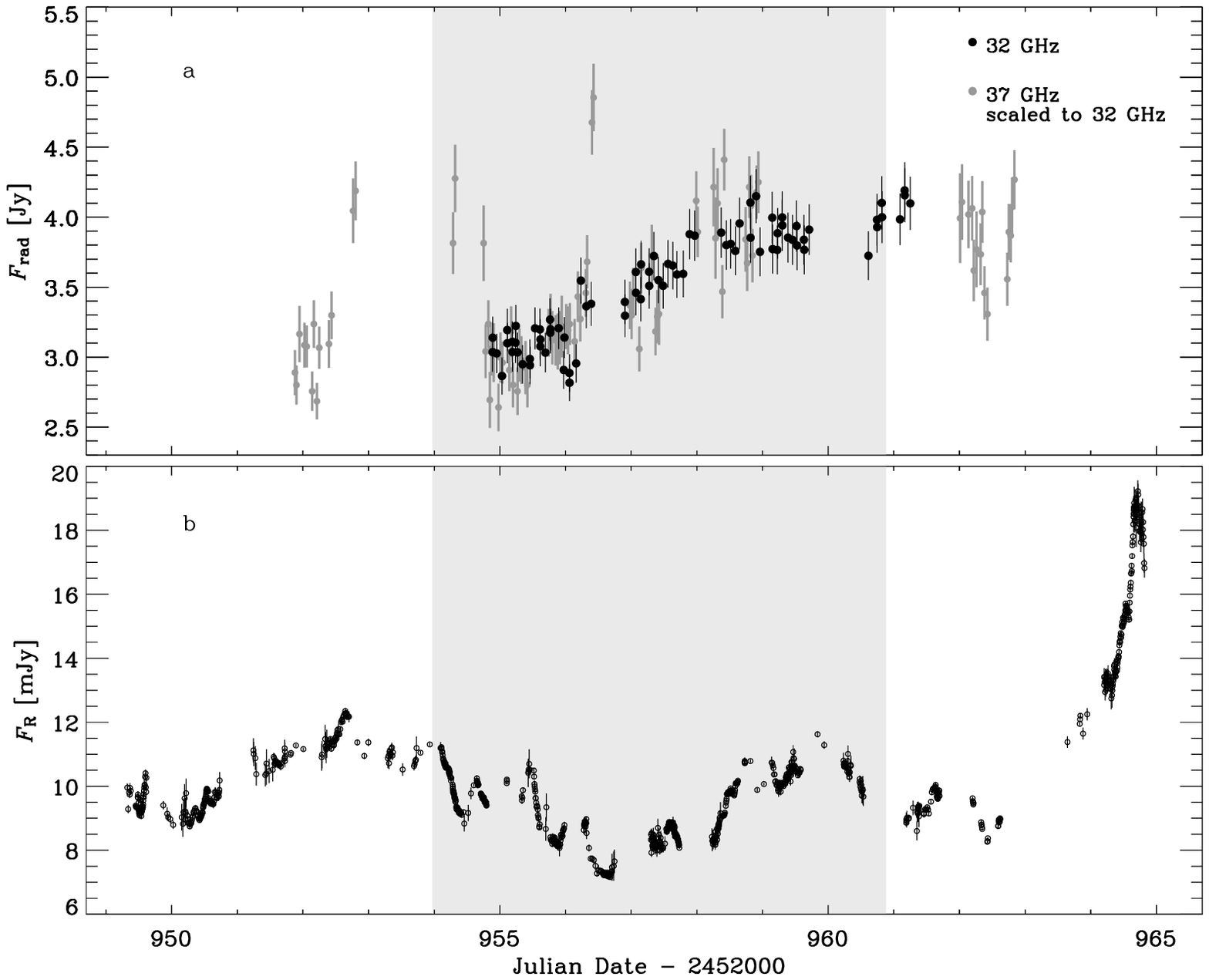}}
\vskip -1.truecm\
\caption{Radio-optical light curve of S5 0716+71 during the core campaign.
         Panel a: 32 GHz radio light curve, and 
         37 GHz radio light curve scaled to the 32 GHz one (scaling factor:
         $<F_\mathrm{32 \, GHz}/F_\mathrm{37 \, GHz}>$=0.89).
         Panel b: $R$-band optical light curve.
         The shaded strip indicates the period of the INTEGRAL pointing.
         The two radio light curves, consistent with each other, exhibit
         significant inter-day variability but low-amplitude IDV. The optical
         light curve displays a weekly modulation of the emission with    
         typical IDV features superimposed.}
\label{radopt}
\end{figure*}

\subsubsection{Observations, data reduction and calibration}
The radio observations of S5 0716+71 at 32 GHz were performed during 
JD = 2452954.889--2452961.251 with the 100-m radio telescope in Effelsberg.
The measurements were carried
out using repeated scans, each of them consisting of 4--8 cross-scans in
azimuth and elevation,
resulting in typical single flux-density exposure times of 120--480 s. 
The antenna temperature on the source was measured through the averaged
pointing-corrected amplitude of the Gaussian-shaped cross-scans and with the 
standard data reduction software of the 100-m telescope.
Regular measurements of the system temperature were used to determine the 
time-dependent fluctuation of the atmospheric opacity, which was used
to correct the measured antenna temperature for each scan.
Several non-variable se\-con\-da\-ry calibrators (within 10$^\circ$--20$^\circ$ of
S5 0716+71) were observed with the same sampling
of the target source, in order to ensure
an accurate calibration of the residual gain fluctuations, 
mainly due to atmospheric effects. 
The total flux-density scale was fixed using the standard 
primary calibrators (e.g. 3C 286, 3C 295, NGC 7027), whose fluxes are given in
Baars et al. (\cite{baars1977}) and Ott et al. (\cite{ott1994}). 
Further details on the observing method and data-reduction scheme can be 
found in Kraus et al. (\cite{kraus2003}) and references therein.

The measurements at 37 GHz were obtained in the period JD = 2452951.878--2452962.838
with the 13.7-m diameter radome-enclosed antenna of the Mets\"ahovi Radio
Observatory. 
The 37-GHz receiver is a dual horn, Dicke-switched receiver with a HEMT
preamplifier, and it is operated at room temperature.
{\it On-on} observations were performed, alternating the source
and the sky in each feed horn, and adopting typical integration
times of 1200--1400 s.
The source DR 21 was used as the primary flux calibrator, whereas 
3C 84 was used as a se\-con\-da\-ry calibrator. 
Errors in the calibrated fluxes were computed by taking both the contribution
of the measurement rms and the uncertainty in the absolute calibration
into account.
More details about the Mets\"ahovi observing system and data reduction can be
found in Ter\"asranta et al. (\cite{terasranta1998}) and re\-fe\-ren\-ces therein.

\subsubsection{The light curves}
Fig.\ \ref{radopt}a shows a superposition of the 32 GHz (black symbols) and 37
GHz (grey symbols) radio light curves.
In spite of the different measurement accuracy and scatter which
characterize the two data sets, a rise of the source flux over the observing 
period, with some shorter-term fluctuations superimposed, is clearly visible
at both wavelengths.

\begin{table}
\caption{Results of the linear fits to the radio light curves 
         at 32 and 37 GHz over different periods during the core-campaign 
         (see text for details). 
         Col\ 1: time interval relevant to the fit;
         Col.\ 2: observing frequency [and
         size of the time bin applied to the light curve in order to make the linear fit
         acceptable at 5\% significance level]; 
         Col.\ 3: linear slope; Col.\
         4: reduced $\chi^2$ of the fit; 
         Col.\ 5: Fit significance.}       
\label{table:slopes}      
\centering          
\begin{tabular}{llccc}    
\hline
\hline       
Time interval    &  $\nu$ (GHz) & Slope              & $\chi _{red}^2$ & Sign.\\
(JD-2452000.00)  &  [bin (h)]   &(Jy/d)              & ~               & (\%)\\
\hline                  
\multicolumn{5}{c}{a}\\
954.889--961.251  & 32          & $+0.19\pm0.01$     & 1.00          & 47.53 \\
951.878--962.838  & 37~[12.4]   & $+0.12\pm0.03$     & 1.38          & 19.04 \\ 
\hline
\multicolumn{5}{c}{b}\\
954.878--958.962  & 32          & $+0.25\pm0.02$     & 0.96          & 55.30 \\
~~                & 37 ~[3.]    & $+0.35\pm0.15$     & 1.22          & 22.71 \\ 
\hline 
\multicolumn{5}{c}{c}\\
954.889--956.391  & 32          & $0.$               & 1.21          & 19.97 \\
~~                & 32          & $+0.08\pm0.06$     & 1.17          & 24.08 \\
956.904--957.968  & 32          & $0.$               & 0.86          & 61.50 \\
~~                & 32          & $+0.35\pm0.13$     & 0.38          & 98.50 \\
958.372--959.711  & 32          & $0.$               & 0.36          & 99.67 \\
~~                & 32          & $-0.01\pm0.09$     & 0.38          & 99.45 \\
960.614--961.251  & 32          & $0.$               & 0.60          & 77.58 \\
\vspace{0.4cm}
~~                & 32          & $+0.48\pm0.28$     & 0.28          & 96.36 \\
951.878--952.437  & 37          & $0.$               & 3.19          & 0.39  \\
~~                & 37          & $0.42\pm0.27$      & 2.26          & 1.61  \\
954.782--955.419  & 37          & $0.$               & 1.13          & 32.26 \\
~~                & 37          & $-0.24\pm0.22$     & 1.12          & 33.46 \\ 
955.743--956.310  & 37          & $0.$               & 0.49          & 92.96 \\
~~                & 37          & $0.42\pm0.27$      & 0.33          & 98.37 \\
956.980--957.423  & 37          & $0.$               & 1.67          & 9.96  \\
~~                & 37          & $-0.10\pm0.37$     & 1.88          & 6.87  \\
957.990--958.938  & 37          & $0.$               & 2.09          & 1.16  \\
~~                & 37          & $0.05\pm0.17$      & 2.24          & 0.79  \\
962.010--962.428  & 37          & $0.$               & 2.23          & 1.73  \\
~~                & 37          & $-1.86\pm0.57$     & 1.09          & 36.32 \\
962.729--962.839  & 37          & $0.$               & 2.23          & 6.30  \\
~~                & 37          & $6.75\pm2.38$      & 0.27          & 84.51 \\
\hline
\multicolumn{5}{c}{d}\\
952.213--952.803  & 37          & $+2.61\pm0.33$     & 3.58          & 46.52 \\
954.286--954.980  & 37          & $-2.14\pm0.30$     & 2.32          & 2.32 \\ 
956.225--956.428  & 37          & $+8.86\pm1.21$     & 2.88          & 3.47 \\ 
\hline
\end{tabular}
\end{table}

A variability test was performed on the two data sets by
checking their consistency with a constant flux level at the 0.1\% 
significance level: the test yielded a negative result, meaning
that the light curves clearly exhibit variability.

The overall increase of the brightness was approximated by a linear slope:
the 32-GHz data can be represented by a linear increase
of the flux of 41\% in 6.4 days, whereas the 
37-GHz data can be described by a flux rise of 39\% in 10.7 days (see Table 
\ref{table:slopes}a).

A consistency check of the two light curves was performed
for the period in which they overlap, namely JD = 2452954.878--2452958.962\ .
In this period, the data at 32 and 37 GHz were characterized by a sampling of
$\sim 12$ and $\sim 15$ data-points per day, and 
average flux uncertainties 
of 4.6\% and  5.0\%, respectively.
The difference in scatter of the two data trains is reflected in the fractional
root-mean-square variability amplitude $F_{\rm var}$ 
(see Vaughan et al. \cite{vaughan2003}, and references therein), 
which is 9.1\% at 32 GHz and 13.9\% at 37 GHz.
However, a Kolmogorov-Smirnov test 
showed that the dif\-fe\-ren\-ces 
between the 32-GHz and the 37-GHz data (rescaled to the 32-GHz 
frequency) are normally distributed at 95\% signi\-fi\-cance level, 
confirming that the measurement errors can well account for the discrepancies
between the two curves. The light curves are hence consistent with each other.

The period of overlap of the two light curves covers the  
inter-day flux variation recorded in the source during the core campaign: 
the rise in flux is consistent at both frequencies with
the linear increase of 35\% in 4.1 days which best fits the 32-GHz data 
(see Table \ref{table:slopes}b).

A search for faster variations was then performed in both light curves for the whole
core-campaign period, during which the sampling of each light curve is such
that it is easy to identify subsets of data 
separated in time by gaps of $\sim 8$ hours.

At 32 GHz, four subsets of light curve can be fitted by a linear trend,
whose slope is significantly variable.
The result of the variability test (made by assuming the slope equal to zero) 
and the best-fit value of the slope are reported, for each subset, 
in Table \ref{table:slopes}c.
Seven subsets of data at 37 GHz can also be approximated by 
linear slopes (see Table \ref{table:slopes}c), with the
exception of seven measurements during  JD = 2452952, 2452954, and 2452956
(see Table \ref{table:slopes}d).
The most extreme deviation was the one recorded during JD = 2452956 
(which yielded a flux difference of $\sim 4 \sigma$ in the distribution 
discussed above),
cor\-respon\-ding to a linear increase of the flux of 42\% in 0.12 days.
The other two strong deviations can be described by linear flux variations of 
33\% in 0.41 days and 20\% in 0.12 days, respectively.

The reliability of the observations can be best estimated by comparing the
measurements taken simultaneously with dif\-fe\-rent telescopes (see e.g. Witzel et
al., \cite{witzel1992}, for a comparison of the 
5-GHz observations of S5 0716+71 performed at the same time with the VLA and the 
Effelsberg telescope in May 1989). 
Therefore, although a careful analysis of the 37-GHz data showed that the 
measurements characterized by extreme deviations were not affected by 
any obvious systematic effect, 
the lack of any counterpart at other radio frequencies did not allow us 
to consider the lower limits to the brightness tem\-pe\-ra\-tu\-res 
inferred from those variations (up to $T_{\mathrm b}\sim 1.5\times 10^{17}$ K)
as a reliable constraint on the variability brightness temperature of the source.

\subsection{Optical data}
\label{sec_opticaldata}

\subsubsection{Observations}
Table \ref{table:telescopes}b (cols.\ 1--4) shows names and locations of the 
nineteen observatories which contributed to the core campaign with $R$-band 
optical data, together with the sizes of the telescopes used and the 
amount of useful observations provided.

It was recommended that the optical observers performed intranight observations of 
S5 0716+71 and of the field re\-fe\-rence stars nos.\ 1 to 8 whose 
magnitudes are given in Villata et al. (\cite{villata1998}).
The choice of the exposure times was left to the observers, who 
found the optimal compromise between high accuracy and good temporal
sampling.

The core-campaign period was characterized by unfavourable weather conditions
in most of the observing sites.
However, the high declination of the source, the long northern nights, and the 
large number of telescopes involved allowed a considerable overlap among 
observations carried out with telescopes located at very
different longitudes.
This led to unprecedented dense monitoring of the source over 
such a long period: 2849 useful observations were performed in 
$\sim 15$ days, with an average rate of $\sim 8$ frames per hour.

\subsubsection{Data reduction, calibration and assembly}

Ten out of nineteen data sets (44\% of the collected data) were homogeneously analysed 
with aperture photometry on de-biased and flat-fielded frames. The remaining
data sets were analysed with different procedures by the observers themselves.

The instrumental magnitudes were then processed in order to obtain the standard
magnitudes of S5 0716+71 and the re\-le\-vant errors.
Due to the variety of characteristics of the telescopes and detectors
involved, as well as of the collected photometric results, not all the 
reference stars suggested to the observers were available for the calibration.
Moreover, because the source was faint during most of the core campaign, the
long exposure times required 
often caused the saturation of re\-fe\-rence stars 1 and 2.
In order to achieve a homogeneous data calibration and minimize  
the instrumental offsets, we defined our calibration sequence
of reference stars common to most of the data sets 
and free of saturation, i.e.\ stars 3, 5 and 6.
The uncertainty in the calibrated magnitudes of S5 0716+71 was estimated, 
for each subset of data taken during the same night and under comparable
observing conditions, as follows: the two stars of the calibration
sequence closest in magnitude to the source were selected, and their magnitude
difference calculated. For each frame, the deviation of this quantity
with respect to its mean was found. The error was taken as the larger 
of this deviation and the standard deviation of the difference over 
the night.

Binning (over time intervals from 5 to 20 minutes) was applied to 
intranight data trains affected by noise and/or low accuracy 
due to either non-optimal sky conditions
or too short integration times.

The calibrated light curves relevant to the various telescopes were then 
assembled according to the procedure described in Villata et al. 
(\cite{villata2002}). Instrumental offsets were computed, night by night,
whenever this was allowed by the temporal overlap between two data sets,
i. e.\  when one of the two datasets had at least two data points belonging to
the time interval covered by the second data set.
Each offset was defined as the mean magnitude difference of the
overlapping parts of the two light curves, the difference being computed   
between pairs of data points separated in 
time by no more than 5 mi\-nu\-tes (see Table \ref{table:telescopes}b, col.\ 5).
The data sets were then corrected by the above offsets.
Data subsets characterized by both lower accuracy and worse temporal sampling
were finally discarded when higher-accuracy and better-sampled data trains
were available in the same time interval.
Data points affected by errors greater than 0.05 mag were not included in the
final light curve, unless no other data were available within 10 minutes.
The average accuracy of the final light curve is of order $\sim 1 \%$.

\subsubsection{The light curve}
The $R$-band core-campaign light curve is displayed in
Fig. \ref{radopt}b. 
Flux densities were derived from magnitudes using the absolute calibrations of Bessel 
(\cite{bessel1979}), and dereddened with 
the ex\-tin\-ction laws of Cardelli et al. (\cite{cardelli1989}), 
under the assumption of a Galactic extinction $A_{\mathrm B}=0.132 \, \rm mag$ 
(provided by NED; from Schlegel et al. \cite{schlegel1998}).

This light curve is characterized by a mean sampling of 20 minutes 
through the whole core-campaign period, and of about 14 minutes during the
INTEGRAL observation. 

The source displayed remarkable variability during the core campaign:
the overall peak-to-peak $R$-band variation was $\Delta F_R=12.1$ mJy 
($\Delta R=1.07$ mag), and the fractional
root mean square variability amplitude $F_{\rm var}=23.3\%$.

A weekly modulation of the emission
with intra-day features superimposed 
characterized the relatively faint state of the source before JD = 2452963.0:
during this period the source brightness varied between 7.2 and 12.4 mJy 
($R=14.17-13.58$), the mean flux was $<F_R>=9.5$ mJy ($<R>=13.87$), and 
$F_{\rm var}= 10.6\%$.
After JD = 2452963.0, 
a brightness rise of 7.8 mJy ($\Delta R=0.57$ mag) occurred in 25.7 hours, and the source
reached 19.2 mJy ($R=13.10$ mag),
the highest level recorded during the core-campaign period.

The IDV features do not generally show well-defined shapes, suggesting that
there might be blending between successive flares. Their linear
parts (in magnitude scale), lasting up to $\sim 3.5$ hours, 
display rising and declining rates not faster than 13\% per hour.
These variation rates are comparable with the steepest slopes 
reported by Wagner et al. (\cite{wagner1996}) ($\sim 10$\% h$^{-1}$) 
and also with the slopes found by Villata et al. 
(\cite{villata2000}) (12\% h$^{-1}$).
Lower-amplitude intra-day variations, on comparable time scales, were 
reported by Ghisellini et al. (\cite{ghisellini1997}), 
Sagar et al. (\cite{sagar1999}) and Nesci et al. (\cite{nesci2002}).
Detailed temporal and spectral analysis of the optical variability 
will be given in forthcoming papers.

\section{INTEGRAL observations and data analysis}
\label{sec_integral}
INTEGRAL (Winkler et al. \cite{winkler2003}) observed S5 0716+71 
from 2003 November 10th, 11:20:04 UT to 2003 November 17th, 09:09:31 UT,
for a total amount of $\sim 539$ ks. 

The observations were strongly affected by the biggest ever recorded 
solar flare (classified as X28), which occurred on November 04, 2003. Because of this
event, SPI (Vedrenne et al. \cite{vedrenne2003}) underwent annealing treatment
during revolutions 132--136 and was not operational for most of the
granted ob\-ser\-ving time.
Of the two detectors of JEM-X (Lund et al. \cite{lund2003}), only JEM-X 2 was used, 
while JEM-X 1 was switched off.
IBIS/ISGRI (Ubertini et al. \cite{ubertini2003}; Lebrun et al. \cite{lebrun2003})
and IBIS/PICsIT (Ubertini et al. \cite{ubertini2003}; Di Cocco et al. \cite{dicocco2003})
were both operational.

The INTEGRAL data were analysed by means of 
the INTEGRAL Offline Scientific Analysis (OSA) software,
whose algorithms are described in Westergaard et al. (\cite{westergaard2003}) 
for JEM-X and Goldwurm et al. (\cite{goldwurm2003}) for IBIS.
Version 4.2 of the OSA sof\-tware was used for the analysis of the IBIS data,
whereas the improved OSA 5.0 was necessary for the JEM-X data analysis.
The PICsIT data could not be analysed due to the failure of the OSA software pipeline.

\subsection{JEM-X} 
\label{sec_jemx}
The JEM-X observation is split into 145 different science windows with an
average exposure of $\sim 3$ ks each. 
The OSA software was used to look for sources within an offset 
angle of $5^{\circ}$ to avoid spurious detections. 
S5 0716+71 was not detected in any of the single science windows, neither 
in the 3-35 keV band nor in any sub-band. 
There was also no other detected source in the field of view.

We combined the images from all the individual science windows into a mosaic 
(corresponding to a total effective exposure of $\sim 381$ ksec) with the 
\textit {varmosaic} routine of the FTOOL package; however, the source remained 
undetected. 

We estimated an upper limit to the flux of S5 0716+71 by means of the 
statistical variance of the mosaic intensity at the position of the source.
The $3\sigma$ upper-limit intensity (cts cm $^{-2}$ s$^{-1}$),
with $\sigma$ being the square root of the variance, 
was multiplied by the ratio of the flux of the Crab Nebula in the JEM-X band 
(Toor \& Seward \cite{toor1974}; N\o rgaard et al. \cite{norgaard1994}) and
the JEM-X intensity obtained from a mosaic image of the Crab. 
Assuming the spectrum of S5 0716+71 to be identical to that of the Crab
($\Gamma=2.1\pm0.03$; see Toor \& Seward \cite{toor1974}), 
we obtained a $3\sigma$ flux upper limit of 
$F_{3-35~\rm {keV}}=6.12\times 10^{-12}\,$ erg cm$^{-2}$ s$^{-1}$
(see Table \ref{table:integral}a),
which is lower than the upper limit derived by Pian et al. 
(\cite{pian2005}) from a shorter (189 ks) observation performed in April 2004.

\begin{table}
\caption{Results of the INTEGRAL analysis: JEM-X and IBIS/ISGRI 3$\sigma$ upper limits to
the flux of S5 0716+71 during November 10--17, 2003.}
\label{table:integral}      
\centering          
\begin{tabular}{lccc}    
\hline
\multicolumn{4}{l}{a.\ JEM-X}\\
\hline
Energy      & Intensity                & Photon                   & Flux \\
range (keV) & (10$^{-5}$ cts/cm$^2$/s) & index $\Gamma^{\,\star}$ & (10$^{-12}$ erg/cm$^{2}$/s)\\
~           &~                         & ~                        & ~     \\
$3-35$      &  $<5.89$                 & $2.1$                    & $<6.12$ \\
~           &~                         & ~                        & ~     \\
\hline                  
\multicolumn{4}{l}{b.\ IBIS/ISGRI }\\ 
\hline
Energy      & Count rate$^{\star\star}$ & Photon                   & Flux \\
range (keV) & (cts/s)                   & index $\Gamma^{\,\star}$ & (10$^{-11}$ erg/cm$^{2}$/s)\\
~           &~                          & ~                        & ~     \\
$15-40$     & $<0.176$                  & $1.6-2.1$                & $<1.34\,-\,<1.41$  \\
$40-100$    & $<0.160$                  & $1.6-2.1$                & $<1.70\,-\,<1.72$  \\
$100-200$   & $<0.140$                  & $1.6-2.1$                & $<5.58\,-\,<5.73$  \\
\hline                                  
\end{tabular}
\begin{list}{}{}
\item[$^{\star}$] Photon index adopted to derive the flux from the count
                  rate, under the assumption that $F(E)\propto E^{1-\Gamma}$ 
\item[$^{\star\star}$] Count rate errors are 1$\sigma$ uncertainties
\end{list}
\end{table}

\subsection{IBIS/ISGRI}
\label{sec_isgri}
The ISGRI-instrument data set consists of 150 science windows,
corresponding to a total effective exposure of 362 ksec.
The standard energy binning
was used for the image deconvolution, and the individual frames were 
accumulated into a co-added mosaic image. 
Several sources, including Mrk 3 and Mrk 6, were reported as
significant detections by the OSA softwa\-re. 

S5 0716+71 was not detected by the source locator at $3\sigma$ significance 
in any energy bin of the mosaic image.
A signal was detected at a significance greater than $1\sigma$
in at least one e\-ner\-gy bin 
between 15 and 200 keV in 54\% of the individual science windows.  
The reliability of these detections was checked by extracting, from each 
science window, the count rate at the sky position of S5 0716+71 and in 
12 background regions located within 1$^{\circ}$ of the source.
A two-sample Kolmogorov-Smirnov test showed that the resulting distributions of 
source and background count rates in each sub-band were consistent with 
them being drawn from the same parent distribution at the 5\% significance level, 
confirming the non-detection of S5 0716+71 at the science-window level also.

In order to estimate an upper limit to the source flux in the ISGRI energy 
band, we extracted, for each sub-band, the value of the statistical variance
of the count rate at the position of the source from the 
mosaic\footnote{Note that the parameter responsible for 
pixel spread in constructing the mosaic was not activated in our ISGRI analysis 
({\it OBS1\_PixSpread}=0), in order to obtain a better estimate of
variance in the mosaic (A. Goldwurm, priv.\ comm.)} of all the science windows.
The corresponding standard deviation ($\sigma_{\rm stat}$) 
was multiplied by the HWHM of the distribution of the significances of the 
mosaic ($\sigma_{\rm syst}$) to take systematic errors into account (A. Goldwurm, priv.\ comm.). 
The $3\sigma$ upper limits ($\sigma=\sigma_{\rm stat} \cdot \sigma_{\rm syst}$) 
to the count rate derived with this procedure are shown in Table \ref{table:integral}b.

The above-mentioned result was crosschecked by e\-sti\-ma\-ting, for each energy sub-band, 
the standard deviation $\sigma_{\rm stat}$ from the statistical variances 
of all the science windows (A. Goldwurm, priv.\ comm.). 
The flux and variance at the nominal position of the source were extracted
from each science window. The mean of these fluxes was then computed by
weigh\-ting it by the inverse of the corresponding variances.
The error in the mean flux, representing $\sigma_{\rm stat}$, was found to 
have the same value as in the case of the mosaic, thus yielding equivalent upper limits.

The count-rate upper limits derived with these methods were converted into fluxes through 
XSPEC by using the results of the spectral analysis of three {\it Beppo}SAX 
observations (Giommi et al. \cite{giommi1999}; Tagliaferri et
al. \cite{tagliaferri2003}) above $\sim 2$ keV, 
where the source spectrum can be well represented by a power law ($\Gamma=1.6-1.96$).
In addition, the Crab Nebula spectrum ($\Gamma=2.1$)
was also considered, for consistency with the assumption
made for the computation of the JEM-X flux.
By varying the assumed photon index in the range $\Gamma=1.6-2.1$,
we obtained the 3$\sigma$ flux upper limits given in Table \ref{table:integral}b.

A comparison of this result with the ISGRI flux of S5~0716+71 derived by Pian
et al. (\cite {pian2005}) in the 30--60 keV range 
from their $\sim 30$\% shorter observation (256 ksec) of April 2004 
was performed by computing the flux in the same e\-ner\-gy bin, under the 
assumption of a spectral index equal to that of the Crab Nebula.
The upper limit obtained in this way was
$F_{30-60~\rm {keV}}=1.06 \times 10^{-11}$ erg cm$^{-2}$ s$^{-1}$, 
a factor of $\sim 3$ below their detection.

\section{The spectral energy distribution (SED)}
\label{sec_sed}

\begin{figure*}
\centering
\includegraphics[width=17cm,height=9.5cm]{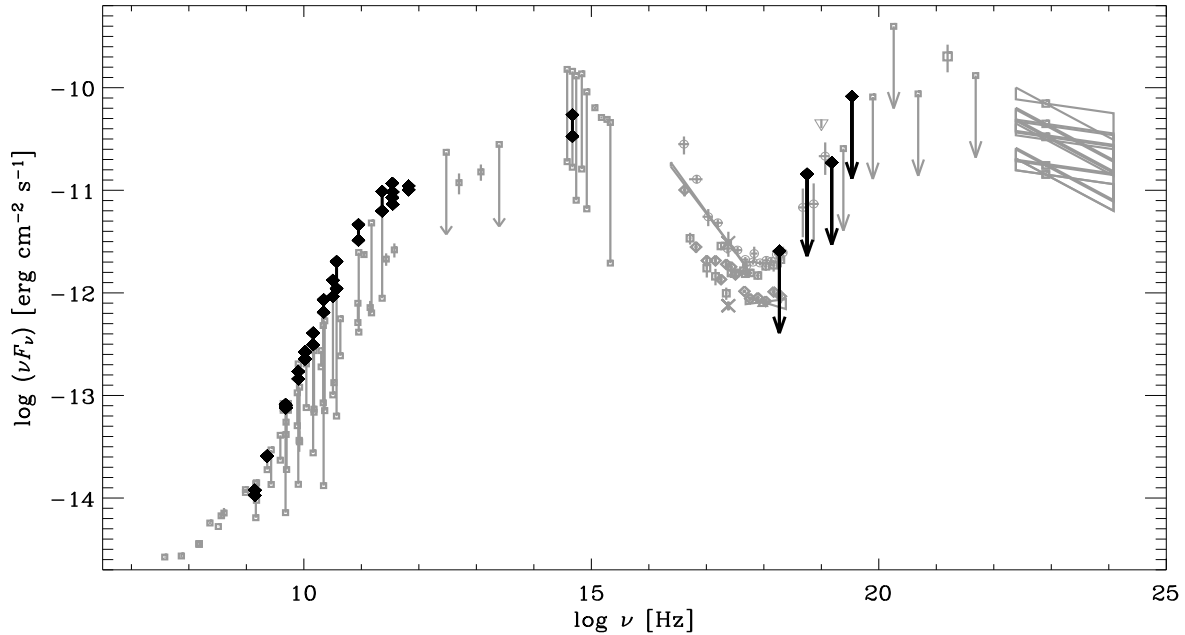} 
\caption{Spectral energy distribution (SED) of S5 0716+71: dark
         diamonds represent data simultaneous with the INTEGRAL pointing, and grey
         symbols represent historical data; variation ranges are indicated
         by vertical bars.
         Simultaneous data are from this work: they show the exceptionally
         bright state of the source in the radio-to-submillimeter domain,
         the moderate level of the optical emission and the upper limits to the
         hard-X--ray brightness.
         Historical data are from:
         RATAN-600, at the wavelengths of 1.38, 2.7, 3.9, 7.7, 13 and 31 cm 
         (Ostorero et al., in prep.);        
         K\"uhr et al. (\cite{kuhr1981}), 
         Waltman et al. (\cite{waltman1981}), Eckart et al. (\cite{eckart1982}), 
         Perley (\cite{perley1982}), Perley et al. (\cite{perley_etal1982}),
         Lawrence et al. (\cite{lawrence1985}), Saikia et al. (\cite{saikia1987}),
         K\"uhr \& Schmidt (\cite{kuhr1990}),
         Moshir et al. (\cite{moshir1990}),
         Hales et al. (\cite{hales1991}),
         Krichbaum et al. (\cite{krichbaum1993}), Gear et al. (\cite{gear1994}),
         Hales et al. (\cite{hales1995}), Douglas et al. (\cite{douglas1996}), 
         Rengelink et al. (\cite{rengelink1997}), Zhang et al. (\cite{zhang1997}),
         Riley et al. (\cite{riley1999}), Cohen et al. (\cite{cohen2002}),
         Raiteri et al. (\cite{raiteri2003}) and references therein,
         at other radio-to-optical frequencies;
         Pian \& Treves (\cite{pian1993}), \&  Ghisellini et al. (\cite{ghisellini1997})
         in the UV band;
         Biermann et al. (\cite{biermann1992}), Comastri et al. (\cite{comastri1997}), 
         Kubo et al. (\cite{kubo1998}),
         Giommi et al. (\cite{giommi1999}), Tagliaferri et al. (\cite{tagliaferri2003}),
         \& Pian et al. (\cite{pian2005}) in the X-ray band;
         McNaron-Brown et al. (\cite{mcnaron1995}), Hartman et al. (\cite{hartman1999}), 
         \& Collmar (\cite{collmar2006})
         in the $\gamma$-ray energy range.}
\label{sed}
\end{figure*}

The broadband monitoring carried out during the INTEGRAL pointing
(see Table \ref{table:telescopes} for a list of the ground-based observing facilities)
allowed us to assemble, for the first time, the SED of S5 0716+71 
with truly simultaneous data spanning more than 10 decades of frequency and
characterized by an exceptional energy sampling.

\begin{figure*}
\centering
\includegraphics[width=17cm,height=9.5cm]{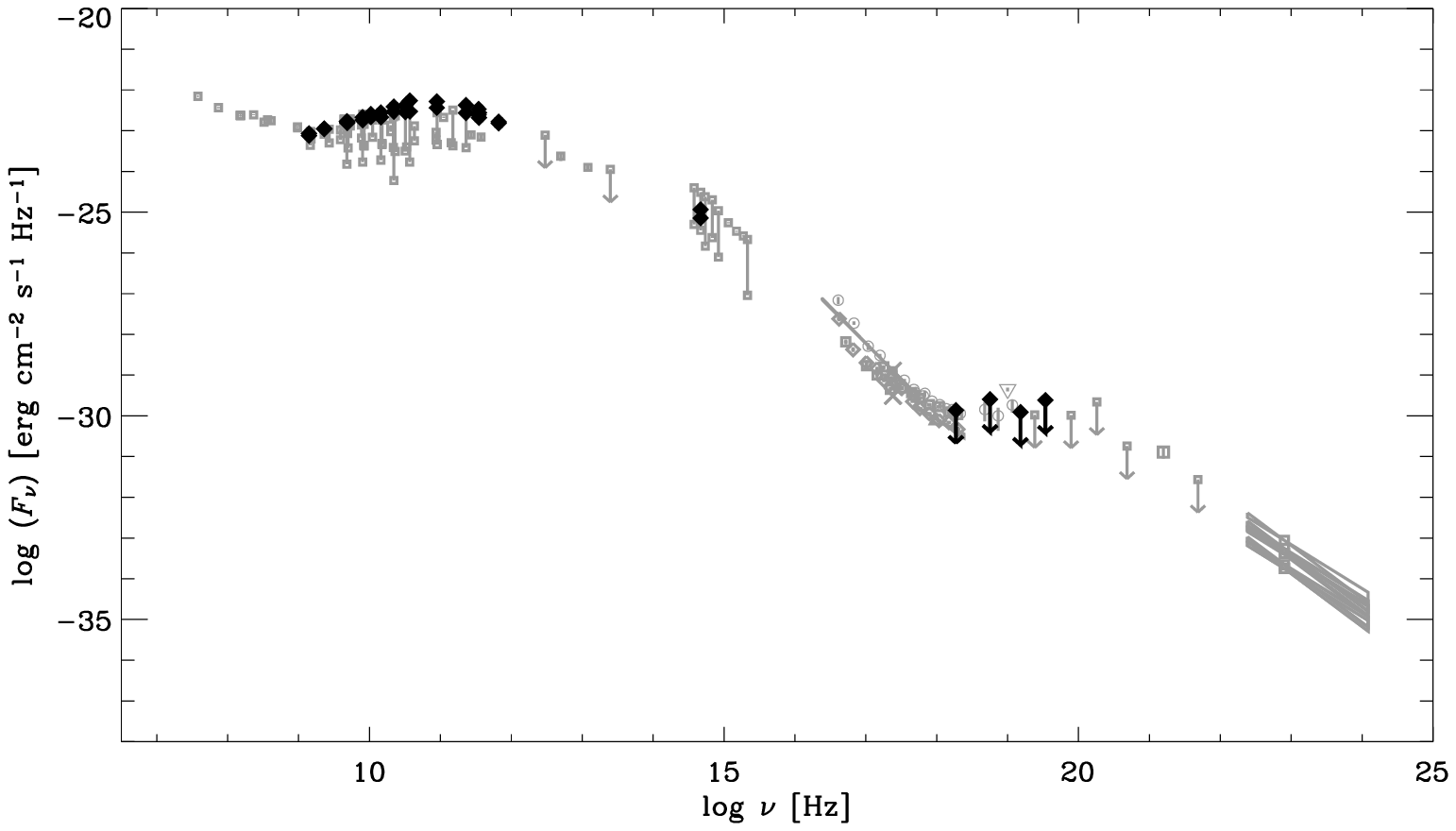}  
\caption{Broadband spectrum of S5 0716+71: dark diamonds represent data 
         simultaneous with the INTEGRAL pointing, and grey 
         symbols represent historical data; variation ranges are indicated 
         by vertical bars.
         See caption of Fig.\ 2 and Sect.\ \ref{sec_sed} for more details.} 
\label{spectrum}
\end{figure*}

The SED of S5 0716+71 is displayed in Fig. \ref{sed}, whereas the 
$F_{\nu}$ vs $\nu$ diagram is shown in Fig. \ref{spectrum} (Online Material).
The flux variation ranges at radio-to-optical frequencies 
recorded during the INTEGRAL pointing period (see Sect. \ref{sec_ground}), 
as well as the upper limits estimated from
the non-detection of the source by INTEGRAL (see Sect. \ref{sec_integral})
are plotted with dark diamonds, and are superposed on the historical data or
variation ranges, drawn with grey symbols\footnote{Previous studies of the 
SED of S5 0716+71 can be found in
Wagner et al. (\cite{wagner1996}), Ghisellini et al. (\cite{ghisellini1997}),
Giommi et al. (\cite{giommi1999}), Ostorero et al. (\cite{ostorero2001}), and 
Tagliaferri et al. (\cite{tagliaferri2003}).}.

The low-energy part of the SED, which is commonly interpreted as due to synchrotron emission, 
shows the unprecedented brightness of the source in the radio-millimeter e\-ner\-gy range
and the moderate emission recorded at optical frequencies. 
The synchrotron peak (representing the most relevant output of the synchrotron
component) is likely to be located around $10^{13}-10^{14}$ Hz.

During the core campaign, the synchrotron spectrum of S5 0716+71 was inverted 
($\alpha >0$, with $F_\nu \propto \nu^{\alpha}$)
in the radio-millimeter wavelength domain, with a turnover frequency close to
90 GHz (see Agudo et al. \cite{agudo2006}).
The spectral index of the partially opaque part of the spectrum was
derived from the simultaneous radio measurements at 5 and 32 GHz (considerably
below the turnover) performed on JD=2452955 and JD=2452961
with the Effelsberg 100-m radio telescope (see Fuhrmann et al., in prep.)
Its value is $\alpha_{5-32}=+0.3$ on JD=2452955, and $\alpha_{5-32}=+0.5$ on JD=2452961:
as the millimeter-flux rose towards the end of the observing period, the
radio spectrum became more inverted.

As usually found in compact radio sources, the spectrum 
at frequencies lower than the turnover was not rising as fast as
the optically thick spectrum of a homogeneous spherical synchrotron source 
($F_{\nu}\propto\nu ^{2.5}$).
Such behaviour is traditionally interpreted as the result of a superposition 
of spectra cha\-racte\-rized by different synchrotron self-absorption (SSA) 
frequencies, which might be produced either by a finite number of homogeneous components
(Kellermann \& Pauliny-Toth \cite{kellermann1969})
or by an inhomogeneous source with gradients in the magnetic field and
particle density (Condon \& Dressel \cite{condon1973}; de Bruyn
\cite{deBruyn1976}; Marscher \cite{marscher1977}).

The variability observed in the radio-millimeter energy range, of which we 
showed an example in the 32 and 37 GHz light curves presented in Sect. 
\ref{sec_radiodata}, hence occurred in a regime in which SSA processes 
might have played an important role.

The IDV study carried out during the fainter radio-to-optical state of 
the source of February 1990 
(Wagner et al. \cite{wagner1990}; Quirrenbach et al. \cite{quirrenbach1991}),
reported a variable radio spectrum which was 
on average optically thin both in the 5.0--8.4 GHz energy range
($\alpha_{5-8.4}=-0.10$)
and in the 1.4--5.0 GHz band 
($\alpha_{1.4-5}=-0.35$) (Wagner et al. \cite{wagner1996}).
The long-term behaviour of the radio spectral indices 
indicates spectra on average inverted in the 4.8--14.5 GHz band 
($\alpha_{4.8-8}=0.72$; $\alpha_{8-14.5}=0.16$) 
during 1985--1992 (Wagner et al. \cite{wagner1996}), as well as in 
the 5.0--15.0 GHz band ($\alpha_{5-15}=0.19$) during 1978--2002 (Raiteri et
al. \cite{raiteri2003}), where also a flatter-when-brighter trend 
and a weak correlation between spectral flattening and ejection of a new component
were recognized by Raiteri et al. (\cite{raiteri2003}) and Bach et
al. (\cite{bach2005}), respectively.
However, no other truly simultaneous spectral information 
on the source spectrum from radio up to millimeter frequencies 
is available to date.

The moderate level of the optical emission recorded du\-ring the bright radio
state is consistent with the absence of correlation, at zero time lag, between major
optical and radio events in the source, as already noticed by Raiteri et
al. (\cite{raiteri2003}) from inspection of the historical (1978--2002) 
radio-optical light curve.

As far as the high-energy part of the SED is concerned, the upper limits 
provided by JEM-X (see Sect. \ref{sec_jemx})
and by IBIS/ISGRI (see Sect. \ref{sec_isgri})
in the two lower-energy bins
are consistent with the levels recorded by ASCA (Kubo et al. \cite{kubo1998})
and {\it Beppo}SAX (Giommi et al. \cite{giommi1999}; Tagliaferri et
al. \cite{tagliaferri2003}) during past observations, and indicate that 
the source was fainter than during the INTEGRAL observation of 
April 2004 (Pian et al. \cite{pian2005}), represented in Fig. \ref{sed} by the grey
triangle at $\sim 41.4$ keV ($\sim 10^{19}$ Hz).  
The two higher-energy IBIS/ISGRI upper limits 
are comparable with the limits derived from the OSSE observations 
(McNaron-Brown et al. \cite{mcnaron1995}).

\section{Discussion and conclusions}

\label{sec_conclusions}
The known high duty cycle of S5 0716+71 was confirmed by the intensive 
simultaneous radio-to-optical monitoring performed during the core campaign.
The source was variable in both the radio and optical bands, 
although the amplitude of IDV
recorded in the radio regime was 
significantly lower than in previous studies.

Let us assume long-term stationary radio emission, and consider a Gaussian
brightness distribution for the variable source region.
If the event which triggers the flux fluctuations
propagates isotropically through the source, 
the flux variability observed in the radio band allows one to estimate the
brightness temperature of the source through the relation:
\begin{equation}
  T_{\mathrm b}=8.47 \times 10^4 \, F_{\nu} \, \left[ \frac{\lambda \, \, 
                d_\mathrm{L}}{t_{\nu}\,(1+z)^2}\right] ^2 
\end{equation}
\noindent
where $F_{\nu}$ is the flux density in Jy, $\lambda$ is the wavelength in cm,
$t_{\nu}$ in years is a function of the mean flux and the flux variation in the
time interval considered according to the relation
\begin{equation}
   t_{\nu}=\frac{<F_{\nu}>}{\Delta F_{\nu}}\,\frac{\Delta t}{(1+z)}\,
\end{equation}
(Wagner et al. \cite{wagner1996}; see also Jones et al. \cite{jones1974}),
$z$ is the source redshift, and 
$d_\mathrm{L}$ is the luminosity distance of the source in Mpc, given by
\begin{equation}      
\label{dL}
 d_\mathrm{L} = \frac{c(1+z)}{H_{0}}
             \int_{0}^{z}\,[(1+z')^2 (1+\Omega_{\mathrm{M}} z')
             -z' (2+z') \Omega_{\mathrm{\Lambda}})]^{-1/2} 
              \mathrm{d}z' 
\end{equation} 
(Carroll et al. \cite{carroll1992}) in the case of a
$\Lambda$-dominated universe ($\Omega_{\mathrm{\Lambda}}=0.7$,
$\Omega_{\mathrm{M}}=0.3$, $\Omega_{\mathrm{k}}=0$; see 
Spergel et al. \cite{spergel2003}). 
Assuming $H_{\rm 0}$=72 km sec$^{-1}$ Mpc$^{-1}$ (Freedman \cite{freedman2001}) and 
a redshift $z>0.3$,
the source luminosity distance would be greater than 1510 Mpc.
The brightness temperature  
derived from the overall increase of the source flux recorded at 32 and 37 GHz 
during the period of overlap of the two-frequency measurements
($\Delta t=4.1$ days) is hence determined to be 
$T_{\mathrm{b}}>(2.1\pm 0.1)\times 10^{14}$~K,  
if the observed flux variations are not affected by any propagation effects.
Higher lower limits to the brightness temperature 
can be derived under less-conservative as\-sumptions: 
a uniform distribution of the source brightness, and/or an anisotropic
propagation of the perturbation which causes the variability would yield 
values up to $T_{\mathrm{b}}>(1.2\pm 0.1)\times 10^{15}$~K.
Moreover, a redshift limit $z>0.52$ would increase 
$T_{\mathrm{b}}$  by a factor of $\sim 3$.
In any case, our brightness temperature exceeds the IC limit of 
$\sim 10^{12}$~K by at least two orders of magnitude.

If relativistic boosting of the radiation is the explanation of this
excessive value, Doppler factors 
$\delta \ga $ $(T_{\mathrm{b}}/10^{12}\,\mathrm{K})^{1/3} (1+z) \ga 8$
would be required in order to lower the intrinsic brightness temperature of
the source below the theoretical limit\footnote{A more accurate value of the 
IC limit, and consequently a better estimate of the Doppler factor derived from 
the brightness temperature, would require detailed knowledge of the
characteristics of the intrinsic spectrum of the source,
including the synchrotron upper cutoff frequency, the self-absorption 
frequency, the radio spectral index, etc. 
(see e.g.\ Kellermann \& Pauliny-Toth \cite{kellermann1969}; 
Blandford \cite{blandford1990}; Readhead \cite{readhead1994}).}.

A recent analysis of the historical data set of VLBI images of S5 0716+71 at
several frequencies performed by Bach et al. (\cite{bach2005}) showed that 
Doppler factors $\sim$ 20--30 are consistent with the proper motion of the 
jet components.
These values, which rule out slower kinematic scenarios of the source jet
presented previously (Witzel et al. \cite{witzel1988}; 
Gabuzda et al. \cite{gabuzda1998}),
could largely account for the above brightness temperature,
provided that the source region responsible for the observed flux va\-ria\-bi\-li\-ty 
had the same kinematic properties as the VLBI jet components.

The source flux evolution recorded in the centimeter--millimeter domain
was always dominated by that of a core of size $\la$0.1 mas.
Confirmation of this evidence was recently provided by the 15-GHz VLBA 
measurements during August 2003 (Kovalev et al. \cite{kovalev2005}) and by
the analysis of space-VLBI (VSOP) observations performed in 2000
(Bach et al. \cite{bach2006}).
The observed variability hence originates from 
either the sub-parsec jet or regions at the base of the jet itself.

If the variability originates from one or more jet components moving according to the
kinematics described by Bach et al. (\cite{bach2005}), 
and characterized by inverted or flat radio spectra 
(possibly becoming steeper as the components move away from the core),
the brightness temperature would only ap\-pa\-ren\-tly exceed the IC limit.
The non-detection of the source by INTEGRAL might hence easily be the 
consequence of the non-occurrence of any Compton catastrophe in the source.
This scenario would also enable us to reconcile the excessive VLBA brightness 
temperature of 
$T_{{\mathrm b},\, z=0}>1.85 \times 10^{13}$ K
derived by Kovalev et al. (\cite{kovalev2005})
with a value lower than the IC limit, assuming a Doppler factor 
$\delta \ga (T_{\mathrm{b}}/10^{12}\,\mathrm{K})\,(1+z) \ga 24$
(note that the VLBA brightness temperature scales by $\delta/(1+z)$).

On the other hand, if the variable core emission is affected by a 
Doppler enhancement different from that of the resolved VLBI jet components, 
the IC-limit violation would be real for Doppler factors 
$\delta \la 8$.
In this case, Compton catastrophes would have 
occurred in the source: the high-energy non-detection would hence provide a
constraint for any model of the source emission taking second-order Compton 
scattering (e.g.\ Bloom \& Marscher \cite{bloom1996}), and hence the 
possibility of Compton catastrophes, into account.

Some alternative explanations for the excessive brightness temperature 
take propagation effects into account.
The correlation between optical brightness and radio spectral index,
as well as the simultaneous change of variability time-scale
observed during the February 1990 campaign 
(Wagner et al. \cite{wagner1990}; Quirrenbach et al. \cite{quirrenbach1991}; 
Wagner et al. \cite{wagner1996}), have been among the strongest
arguments against the extrinsic origin of radio IDV in S5 0716+71 in 
the past decade. 
However, the radio and optical intra-day variability observed during 
our campaign, when the source was in a brighter radio-to-optical state 
and had an inverted spectrum up to millimeter frequencies,
do not appear to be obviously correlated. 
This might be the consequence of optical-depth effects 
(which can modify the shape and amplitude of flares) on radio emission
intrinsically correlated with that in the optical band; alternatively,  
the radio and optical radiation observed might have come from non-cospatial 
components characterized by different sizes.
In the absence of correlated radio-optical variations, the possibility of 
a contribution of ISS to the observed radio 
variability cannot be completely ruled out, implying that
the brightness temperature inferred from variability
is not representative of the photon density of the source;
in particular, the violation of the IC limit would again be only apparent
in an ISS-driven variability scenario.

It is very unlikely that this variability can be explained by scintillation alone.
In fact, ISS is not a very efficient mechanism in the 
32-37 GHz (8-9 mm) regime, to which our brightness temperature refers.
Moreover, the simultaneous observations at 86 GHz (3 mm), where ISS is less 
efficient than at 32-37 GHz, showed that ISS can be definitely 
ruled out as an explanation for the observed (comparable)
variability at these frequencies and for the corresponding 
excessive brightness temperatures 
(Agudo et al. \cite{agudo2006}).
Therefore, if ISS contributed to the observed 32-37 GHz flux evolution,
in all likelihood it was not the dominant variability mechanism.

In conclusion, violations of the brightness-temperature IC limit were inferred
from the radio variability observed at 32-37 GHz. 
If the recorded radio emission is either intrinsic and beamed with Doppler 
factors $\delta \ga 8$, 
or strongly affected by ISS, the violation of the IC limit would be apparent, 
and readily consistent with the non-detection of the source by INTEGRAL in the
X--$\gamma$-ray regime.
Intrinsic flux variations in the presence of lesser beaming effects would instead 
imply a real violation of the theoretical limit; in this case, 
the non-detection of corresponding $\gamma$-ray avalanches provides 
a strong constraint for modelling of the Compton catastrophe in this source.
At any rate, our unprecedented simultaneous broad-band measurements 
will help to define the parameter space of SSC emission models in  
detailed studies of the multifrequency properties of S5 0716+71.

\begin{acknowledgements}
        We acknowledge an anonymous referee for a careful reading of the manuscript
        and for helpful advice.
        We acknow\-ledge support by BMBF, through its agency DLR for the project
        50OR0303 (S. Wagner).
        We acknowledge EC funding under contract HPRN-CT-2002-00321 (ENIGMA).
        V.\ M.\ Larionov and V.\ Hagen-Thorn 
        acknowledge support from the Russian
        Foundation for Basic Research under grant 05-02-17562.
        S.\ Britzen acknowledges support by the Claussen-Simon Stiftung.
        A.\ S.\ Hojaev acknowledges NSC, Taiwan for an invitation and support 
        as a Visiting Expert at IoA of NCU.

        This research has made use of:
        optical observations on the WHT, La Palma, made with the ULTRACAM photometer: 
        we acknow\-ledge the support of V.\ S.\ Dhillon (University of Sheffield, UK) and
        T.\ R.\ Marsh (University of Warwick, UK);
        data from the University of Michigan Radio Astronomy Observatory,
        which is supported by the National Science Foundation and by funds from 
        the University of Michigan;
        data from the Westerbork Synthesis Radio Telescope, operated by ASTRON with
        financial support from the Netherlands Organisation for Scientific
        research (NWO);
        the NASA/IPAC Extragalactic Database (NED), which is operated by            
        the Jet Propulsion Laboratory, California Institute of Technology, 
        under contract with the National Aeronautics and Space Administration;
	the CATS database (Verkhodanov et al. \cite{verkhodanov1997})
        of the Special Astrophysical Observatory.

        We are grateful to A. P. Marscher, A. Mastichiadis and J.\ G.\ Kirk for stimulating
        discussions on the problems of the brightness tem\-pe\-ra\-ture and IC
        catastrophes, to the INTEGRAL Team for useful advice on the
        INTEGRAL data analysis, and to Y.\ Y.\ Kovalev for discussing the paper.

\end{acknowledgements}

\Online

\end{document}